\documentclass[useAMS,usenatbib]{mn2e}

\def\Msol{\hbox{M$_\odot$}}
\def\Zsun{\hbox{Z$_\odot$}}

\def\kms{\hbox{km$\,$s$^{-1}$}}
\def\cmt{\hbox{cm$^{-3}$}}
\def\Apix{\hbox{\AA$\,$pix$^{-1}$}}
\def\one{\,{\sc i}}             
\def\two{\,{\sc ii}}

\newcommand\fsec{\hbox{$.\!\!^{\rm s}$}}

\usepackage{amsmath}
\usepackage{amssymb}
\usepackage{mathrsfs}
\usepackage{color}
\usepackage{upgreek}
\usepackage{graphicx}

\defcitealias{hunter94b}{H94b}
\title[Ionized gas in NGC 1140]{Ionized gas in the starburst core and halo of NGC 1140\thanks{Based on observations made with the WIYN SparsePak and DensePak instruments.}}
\author[M.S.\ Westmoquette et al.] {M.\ S.\ Westmoquette$^1$\thanks{E-mail:msw@star.ucl.ac.uk}, J.\ S.\ Gallagher III$^2$, L.\ de Poitiers$^1$ \\
$^1$Department of Physics and Astronomy, University College London, Gower Street, London, WC1E 6BT\\
$^2$Department of Astronomy, University of Wisconsin-Madison, 5534 Sterling, 475 North Charter St., Madison WI 53706, USA\\
}
\date{Accepted 2009 December 17. Received 2009 December 3; in original form 2009 October 15}
\pagerange{\pageref{firstpage}--\pageref{lastpage}}
\pubyear{2009}
\begin{document}
\maketitle
\label{firstpage}
\begin{abstract}
We present deep WIYN H$\alpha$ SparsePak and DensePak spatially-resolved optical spectroscopy of the dwarf irregular starburst galaxy NGC 1140. The different spatial resolutions and coverage of the two sets of observations have allowed us to investigate the properties and kinematics of the warm ionized gas within both the central regions of the galaxy and the inner halo. We find that the position angle of the H$\alpha$ rotation axis for the main body of the galaxy is consistent with the H\one\ rotation axis at PA = $39^{\circ}$, but that the ionized gas in the central $20\times 20$~arcsecs ($\sim$$2\times 2$~kpc) is kinematically decoupled from the rest of the system, and rotates at a PA approximately perpendicular to that of the main body of the galaxy at $+40^{\circ}$. We find no evidence of coherent large-scale galactic outflows. Instead multiple narrow emission line components seen within a radius of $\sim$1--1.5~kpc, and high [S\two]/H$\alpha$ ratios found beyond $\sim$2~kpc implying a strong contribution from shocks, suggest that the intense star formation is driving material outwards from the main star forming zone in the form of a series of interacting superbubbles/shells.

A broad component (100\,$\lesssim$\,FWHM\,$\lesssim$\,230~\kms) to the H$\alpha$ line is identified throughout galaxy disk out to $>$2~kpc. Based on recent work looking at the origins of this component, we conclude that it is produced in turbulent mixing layers on the surfaces of cool gas knots embedded within the ISM, set up by the impact of the ionizing radiation and fast-flowing winds from young massive star clusters. Our data suggest a physical limit to the radius where the broad emission line component is significant, and we propose that this limit marks a significant transition point in the development of the galactic outflow, where turbulent motion becomes less dominant. This mirrors what has recently been found in another similar irregular starburst galaxy NGC 1569.
\end{abstract}


\section{Introduction}\label{intro}

Low-metallicity, starbursting, dwarf galaxies are a particularly interesting class of galaxy for a number of reasons: (1) their low metallicities make them excellent analogues to intensely star forming galaxies at high-redshift, which in hierarchical galaxy formation models are thought to be the building blocks of present-day massive galaxies, and (2) the starburst event can have much more of a destructive effect on the galaxy than for more massive systems; their low gravitational potentials may also mean that for those with flattened morphologies, outflowing material is much more likely to escape altogether \citep{larson74, maclow99} \citep[note the reverse may be true for those with more irregular morphologies;][]{silich01}. Although the ejection of the ISM (including the freshly chemically enriched material) has clear consequences for the future evolution of the system in terms of its dynamics and star formation rate \citep{dekel86}, and for the enrichment of the intergalactic medium \citep{aguirre01a, cen05}, some studies suggest that ejection of hot gas through bubble blow-out may not be as efficient as first thought \citep{deyoung94, martin98}. It is therefore important to study such systems to understand how gas is removed and what effects this has in the evolution of the galaxy. 

NGC 1140 is a nearby \citep[20~Mpc;][]{moll07}, low metallicity (LMC-like, 0.4~\Zsun; \citealt{izotov04}; \citealt*{nagao06}; \citealt{moll07}) blue compact dwarf (BCD) galaxy with a total star formation rate of $0.7\pm 0.3$~\Msol~yr$^{-1}$ (Hunter, van Woerden \& Gallagher 1994, hereafter \citetalias{hunter94b}; \citealt{moll07}). The current star formation is dominated by an extremely H$\alpha$-luminous central concentration that is known to host a number of young massive clusters (YMCs) with ages $<$5~Myr and masses $>$$10^5$~\Msol\ (\citealt*{hunter94a}; \citealt{moll07}); this extreme youth is supported by the detection of strong WR star signatures \citep{moll07}. It is thought that these clusters are the most recent products of a galaxy-wide starburst induced by a merger or interaction with a low luminosity, gas-rich companion \citepalias{hunter94b}: the optical colours \citepalias{hunter94b}, broad-band photometry \citep{cidfernandes03}, and photometric star cluster age-dating \citep{degrijs04} all suggest extensive star formation throughout the galaxy in the last 1~Gyr, and particularly in the last 20~Myr. 

\begin{figure*}
\centering
\includegraphics[width=\textwidth]{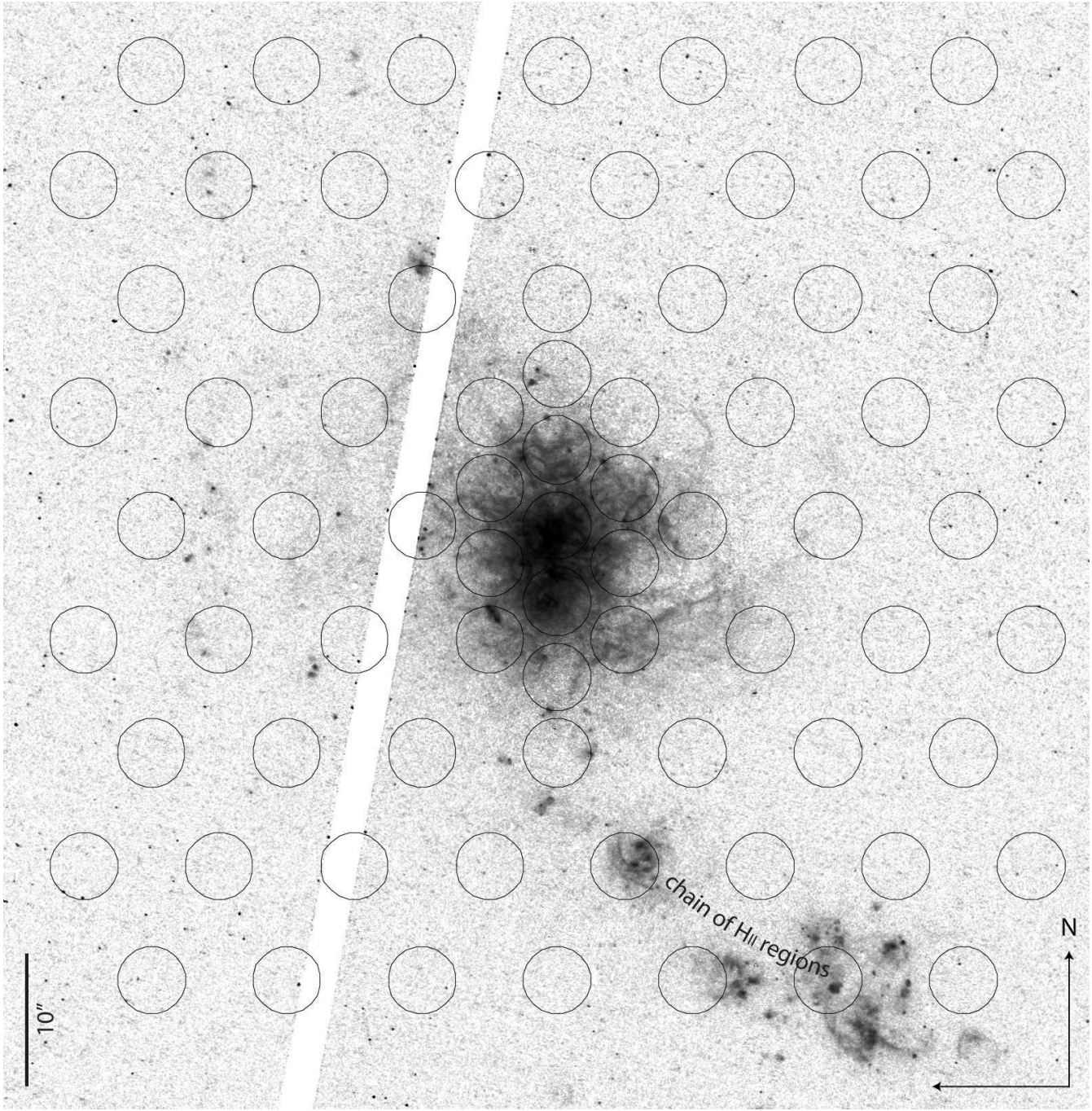}
\caption{\textit{HST}/ACS-WFC continuum subtracted H$\alpha$ image of NGC 1140 (inverse log scaled) with SparsePak footprint overlaid.}
\label{fig:SP_finder}
\end{figure*}

The misalignment of the galaxy's structural components are consistent with the merger scenario. On large scales ($\lesssim$40~kpc), the H\one\ gas is elongated along a position angle (PA) of $-51^{\circ}$ (although strongly warped on the south-eastern side). In the inner regions (central $\sim$10~kpc), the H\one\ gas is concentrated along an approximately perpendicularly oriented ridge (PA $\approx$ $+22^{\circ}$). \textbf{To first order this neutral gas all rotates in almost solid-body rotation, with the line of nodes at the PA of $-51^{\circ}$ (i.e.\ H\one\ rotation axis PA = +$39^{\circ}$)}, and exhibits a high velocity dispersion indicating that the system is not relaxed \citepalias{hunter94b}. In the $V$-band however, the morphology of this inner region is significantly different, being roughly rectangular in shape with a PA of $0^{\circ}$ \citepalias{hunter94b}. At the south-west end, at a distance of $\sim$2.5~kpc, the rectangle has a hook that extends towards the west, traced by a chain of H\two\ regions (see Fig.~\ref{fig:SP_finder}). This chain is aligned with the PA $\approx$ $+22^{\circ}$ H\one\ ridge, and it is these H\two\ regions that are coincident with the peak of the H\one\ emission \citepalias{hunter94b}, not the central H$\alpha$ concentration as might be expected. Further out at fainter $V$-band surface brightness levels, the galaxy appears elongated towards the north-west and south-east along a PA $\approx$ $-15^{\circ}$. Further out still, a number of low surface brightness shells are observed that may be linked with either past star-formation events or the merger/interaction \citepalias{hunter94b}.

\textit{HST} imaging (Fig.~\ref{fig:SP_finder}) shows that the central H$\alpha$ concentration extends into a network of complex shells, filaments and bubbles that fill the ISM of the galaxy out to a distance of $\sim$2~kpc. High [S\two]/H$\alpha$ line ratios (\citetalias{hunter94b}; \citealt{burgh06}) and strong [Fe\two] emission \citep{degrijs04} are both indicators that shocks play an important role in the halo. Thus the presence of multiple YMCs, the high H$\alpha$ luminosity, strong evidence of shocks, and this ``froth'' of structure in the ionized gas all suggest strong starburst-driven feedback is taking place, although to date no kinematic evidence has been found for a coherent, galaxy-wide outflow.


With all this in mind, we have obtained deep spatially-resolved optical spectra of NGC 1140 from two instruments with different spatial resolutions and coverage, to investigate the properties and kinematics of the warm ionized gas within both the central regions of the galaxy and the inner halo.

In this work, we adopt a systemic velocity for NGC 1140 of 1475~\kms\ giving a distance of 20~Mpc \citep{moll07}, meaning $1''\sim 97$~pc.

\section{Observations and Data Reduction} \label{sect:data}

We observed NGC 1140 on two separate occasions with the WIYN\footnote{The WIYN Observatory is a joint facility of the University of Wisconsin-Madison, Indiana University, Yale University, and the National Optical Astronomy Observatories.} SparsePak and DensePak instruments. These two data sets, providing observations at complimentary spatial resolutions, have allowed us to examine the kinematics and nebular properties of the galaxy's disk, halo and nuclear regions in some detail.

\subsection{SparsePak}
Observations of NGC 1140 were obtained with the SparsePak instrument \citep{bershady04} on the WIYN 3.5-m telescope. SparsePak is a ``formatted field unit'' similar in design to a traditional IFU, except that its 82 fibers are arranged in a sparsely packed grid, with a small, nearly-integral core \citep{bershady04}. SparsePak was designed to maximise throughput and spectral resolution at the expense of spatial coverage/resolution, and as such has a total light throughput of $\sim$90 per cent longwards of 500~nm. Each fibre has a diameter of 500~$\upmu$m, corresponding to $4\farcs69$ on the sky; the formatted field has approximate dimensions of $72\times 71.3$~arcsecs, including seven sky fibres located on the north and west side of the main array separated by $\sim$$25''$. The mapping order of fibers between telescope and spectrograph focal planes was purposefully designed in a fairly randomised fashion in order to distribute sky fibres evenly along the slit and minimise the effects of spectrograph vignetting on the summed sky spectrum. SparsePak is connected to the Hydra bench-mounted echelle spectrograph which uses a T2KC $2048\times 2048$ CCD detector.

We obtained observations of NGC 1140 centred on the H$\alpha$ peak at $\alpha$=$2^{\rm h} 54^{\rm m} 33\fsec43$, $\delta$=$-10^{\circ} 01' 39\farcs3$ (J2000) during the period 13--16th December 2004 with a total exposure time of 3$\times$1800~secs. The position of the SparsePak footprint is shown in Fig.~\ref{fig:SP_finder}, overlaid on an \textit{HST}/ACS-WFC continuum subtracted F658N image (Prop ID: 9892, PI: Jansen). Using an order 8 grating with an angle of 63.25$^{\circ}$ (giving a spectral coverage of 6450--6865~\AA\ and dispersion of 0.20~\Apix), we were able to cover the nebular emission lines of H$\alpha$, [N\two]$\lambda\lambda 6548,6583$, and [S\two]$\lambda\lambda 6716,6731$. A number of bias frames, flat-fields and arc calibration exposures were also taken together with the science frames.

\begin{figure}
\centering
\includegraphics[width=0.5\textwidth]{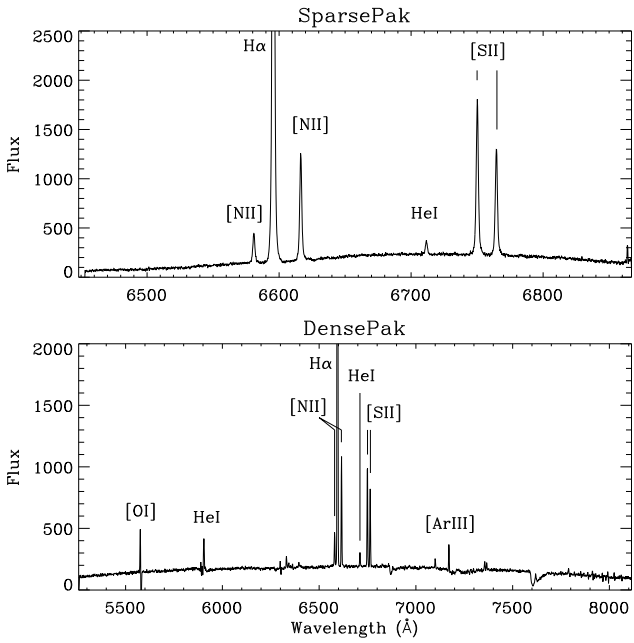}
\caption{Example SparsePak and DensePak spectra showing the full spectral range of each observation (in arbitrary flux units). The identified nebular lines are labelled. A number of residuals remain in the DensePak spectrum as a result of an imperfect sky subtraction.}
\label{fig:spec_labelled}
\end{figure}

Basic reduction was achieved using the {\sc ccdproc} task within the {\sc noao} {\sc iraf} package. Instrument-specific reduction was then achieved using the {\sc hydra} tasks also within the {\sc noao} package. The first step was to run {\sc apfind} on the flat-field exposure to automatically detect the individual spectra on the CCD frame. The output of this task is an aperture identification table which can be used for each science frame to extract out the individual spectra. The task {\sc dohydra} was then used to perform the bias subtraction, flat-fielding and wavelength calibration. The datafile at this stage contained 82 reduced and wavelength calibrated spectra. A representative sky spectrum was created by averaging the seven sky fibre spectra, and sky-subtraction was achieved by re-running {\sc dohydra} with the sky-subtraction option switched on. Cosmic-rays were cleaned from the data using {\sc lacosmic} \citep{vandokkum01}, before final combination of the individual frames was achieved using {\sc imcombine}. The final datafile now contained 82 reduced, wavelength calibrated and sky-subtracted spectra. An example reduced and labelled spectrum is shown in the top panel of Fig.~\ref {fig:spec_labelled}.

In order to determine an accurate measurement of the instrumental contribution to the spectrum broadening, we selected high S/N spectral lines from a wavelength calibrated arc exposure that were close to the H$\alpha$ line in wavelength, and sufficiently isolated to avoid blends. After fitting these lines with Gaussians in all 82 apertures, we find the average instrumental width is $0.7\pm0.02$~\AA\ = $31.4\pm 0.5$~\kms.

\subsubsection{Emission line fitting} \label{sect:line_fitting}
The S/N and spectral resolution of the data have allowed us to accurately quantify the line profile shapes of the emission lines. In the central regions where the S/N is highest, we find the emission lines to be composed of two bright, narrow components (FWHM $\sim$ 20--100~\kms; \textbf{corrected for instrumental broadening}) overlaid on a fainter broad component (FWHM $\sim$ 100--230~\kms; \textbf{also corrected for instrumental effects}). In some spaxels further out, we see only the two narrow components. Two split narrow components are indicative of expanding gas motions.

To quantify the emission line profile shapes, we fitted model Gaussian profiles to each line detected in each spaxel of the SparsePak field using a customised IDL-based $\chi^2$ fitting routine called \textsc{pan} \citep[Peak ANalysis;][]{dimeo}. A detailed description of the program and the customisations we have made to it are given in \citet{westm07a}. As mentioned above, in many cases we found that additional Gaussian components were needed to satisfactorily fit the integrated line profile; the number of components fitted to each line was determined using a combination of visual inspection and the $\chi^{2}$ fit statistic output by \textsc{pan}. For the multi-component cases, we specified a number of rules to identify the different line components. This consistent approach helped both in the minimisation of the fit and in the subsequent analysis. For double-components, the second component (hereafter C2) was sometimes required to fit a broad underlying profile, and in others to fit the fainter component of a split line. Where three components were identified, the broadest component was assigned to component 2 (C2), and after that, the brightest to component 1 (C1) and the tertiary component to C3. We note here that in the double-component cases where a broad component was not detected, the secondary narrow component can be thought of as equivalent to C3 in the triple-component cases (where C2 represents the broad component). \textbf{In all cases the individual Gaussian models were constrained to have a width greater than the aforementioned instrumental width.}

\begin{figure*}
\centering
\begin{minipage}{7cm}
\includegraphics[width=7cm]{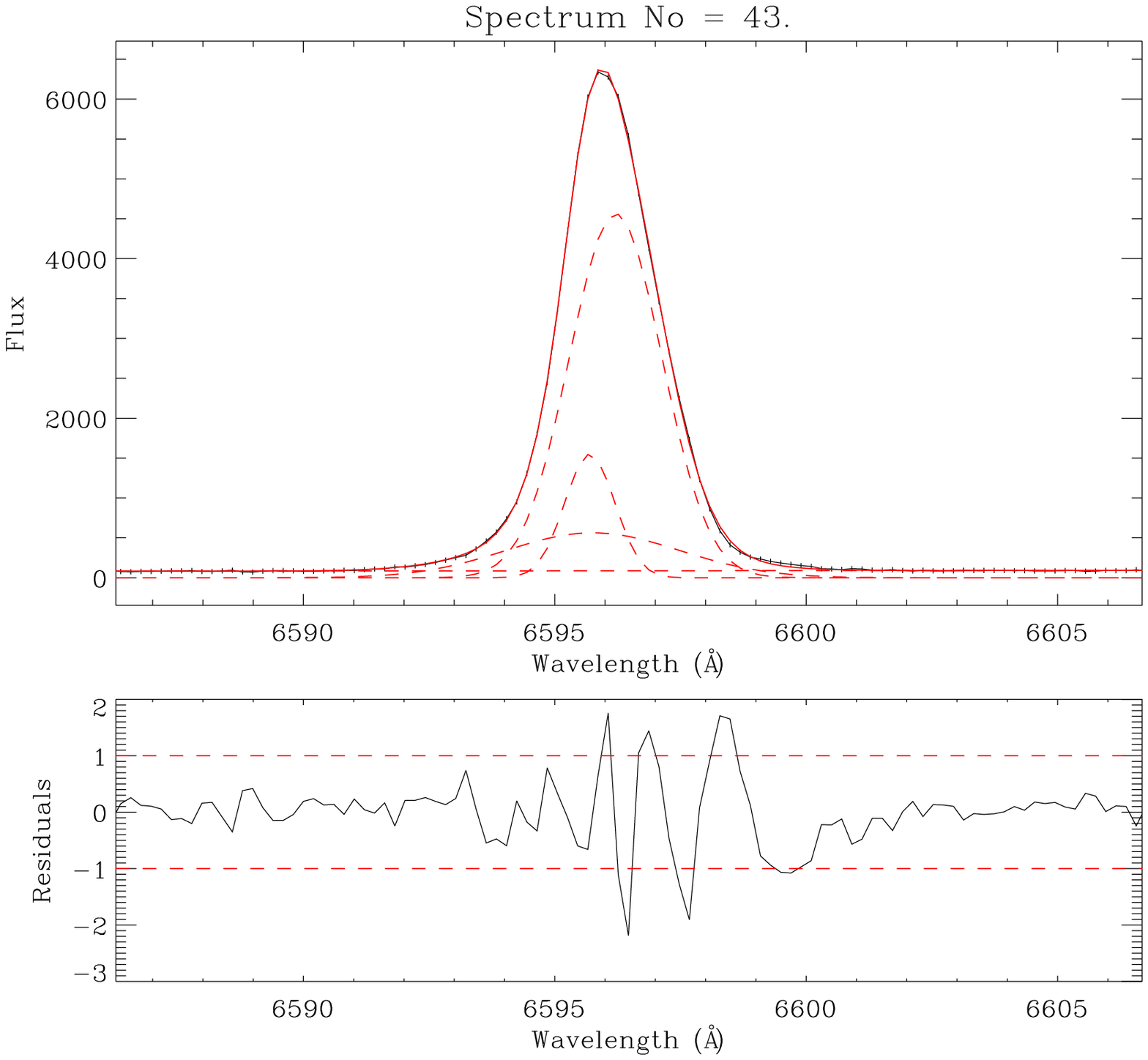}
\end{minipage}
\hspace{0.8cm}
\begin{minipage}{7cm}
\includegraphics[width=7cm]{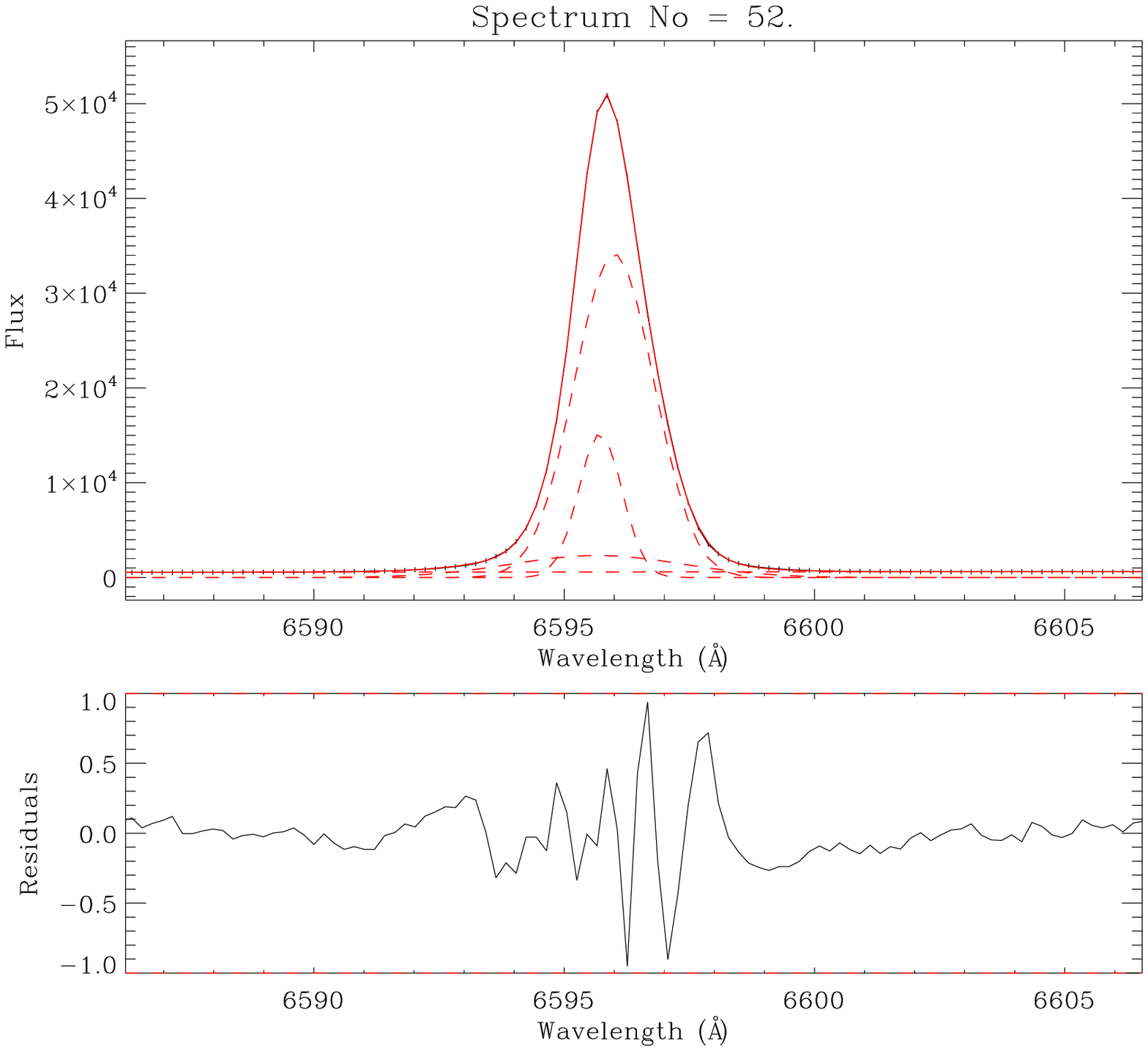}
\end{minipage}
\begin{minipage}{7cm}
\vspace{0.5cm}
\includegraphics[width=7cm]{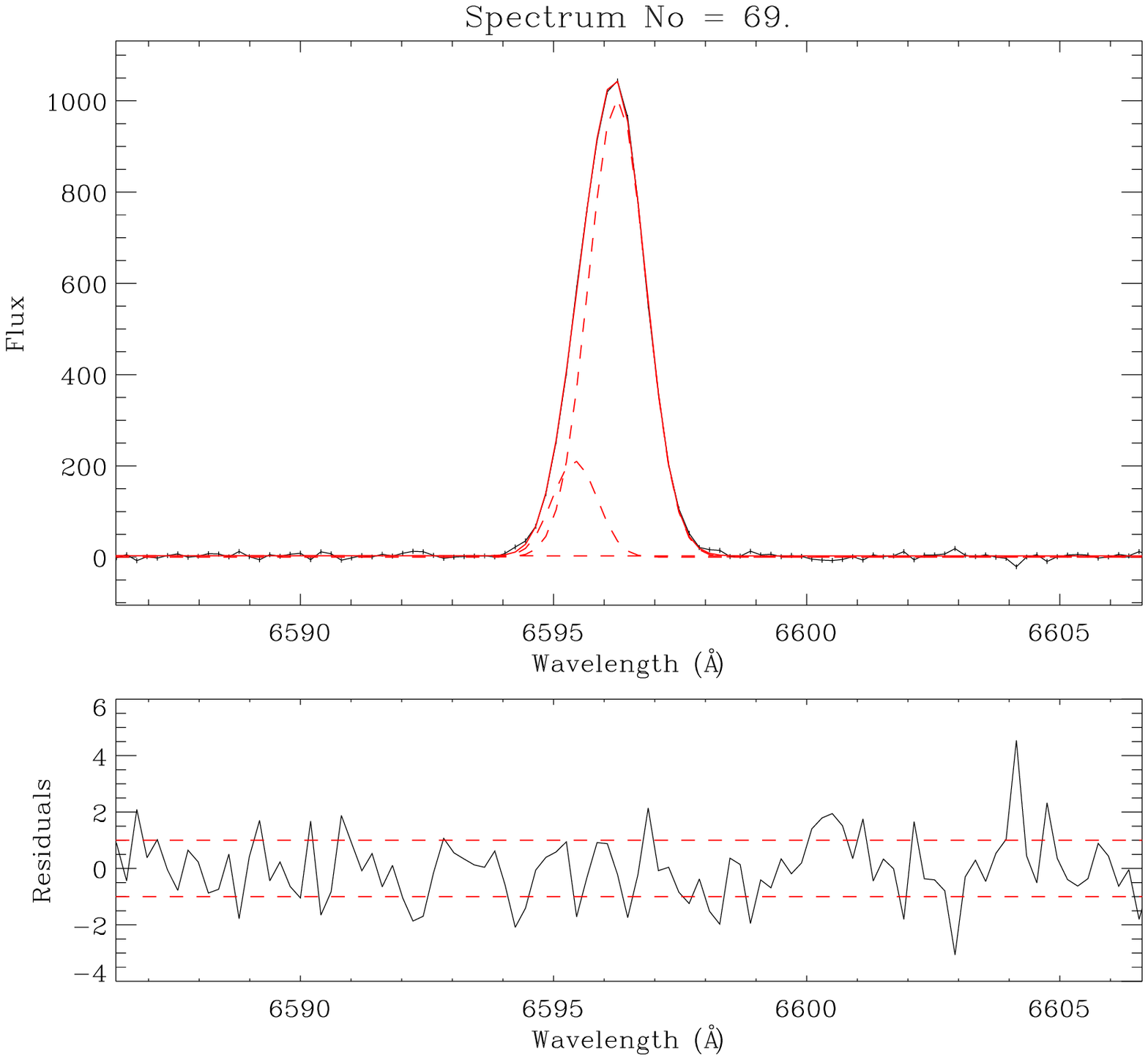}
\end{minipage}
\hspace{0.8cm}
\begin{minipage}{7cm}
\vspace{0.5cm}
\includegraphics[width=7cm]{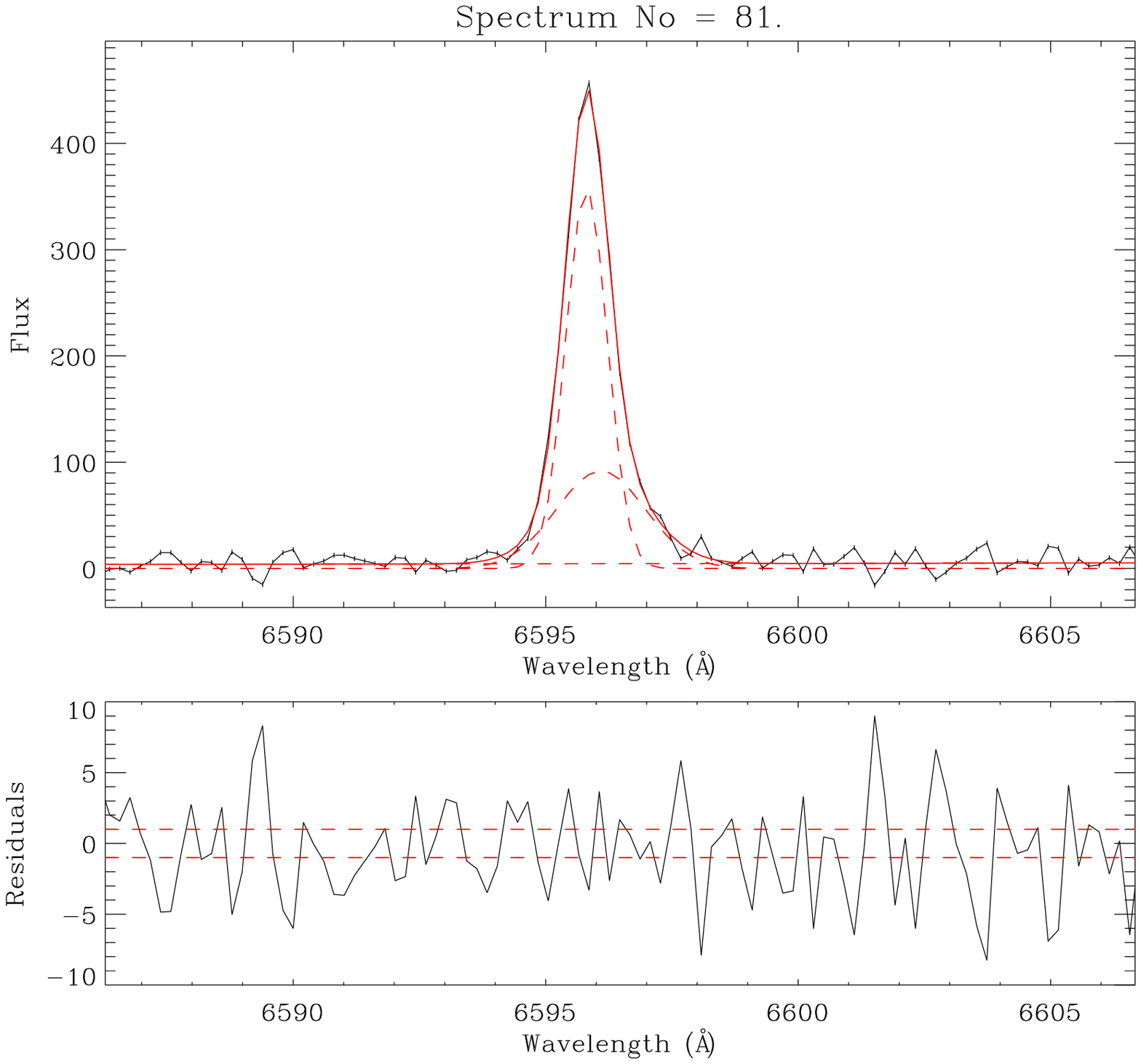}
\end{minipage}
\caption{Example H$\alpha$ line profiles chosen to represent the main types of profile shapes observed over the field. Observed data is shown by a solid black line, individual Gaussian fits by dashed red lines (including the straight-line continuum level fit), and the summed model profile in solid red. Flux units are arbitrary but on the same scale. Below each spectrum is the residual plot.}
\label{fig:sp_eg_fits}
\end{figure*}

A number of example H$\alpha$ line profiles from various spaxels are shown in Fig.~\ref{fig:sp_eg_fits}, together with the individual Gaussian fits required to model the integrated line-shapes. These were compiled to show the variation in profile shapes found.

\subsection{DensePak}
DensePak \citep{barden98} was a small fibre-fed integral field array that was usually attached at the Nasmyth focus of the WIYN telescope\footnote{DensePak was decommissioned in 2008.}. During our run DensePak was instead attached to the f13.7 modified Cassegrain port, thus changing the detector plate-scale from the documented value. The array has 91 fibres, each with a diameter of 300~$\mu$m ($1.3''$ on the sky at the Cassegrain focus). The fibre-to-fibre spacing is 400~$\mu$m making the overall dimensions of the array $12.4\times 19.8$~arcsecs. Four additional fibres are offset by $\sim$$30''$ from the array centre and serve as dedicated sky fibres. The arrangement of the DensePak fibre array on the sky, including the sky fibres, is shown in \citet[][note the dimensions given in this reference apply to DensePak at the Nasmyth focus]{sawyer97}. At the time of observation, there were 8 damaged and unusable fibres in the main array making a usable total of 83. DensePak's fibre bundle was reformatted into a `pseudo-slit' to feed the Hydra bench-mounted echelle spectrograph, just as for SparsePak.

\begin{figure}
\centering
\includegraphics[width=0.48\textwidth]{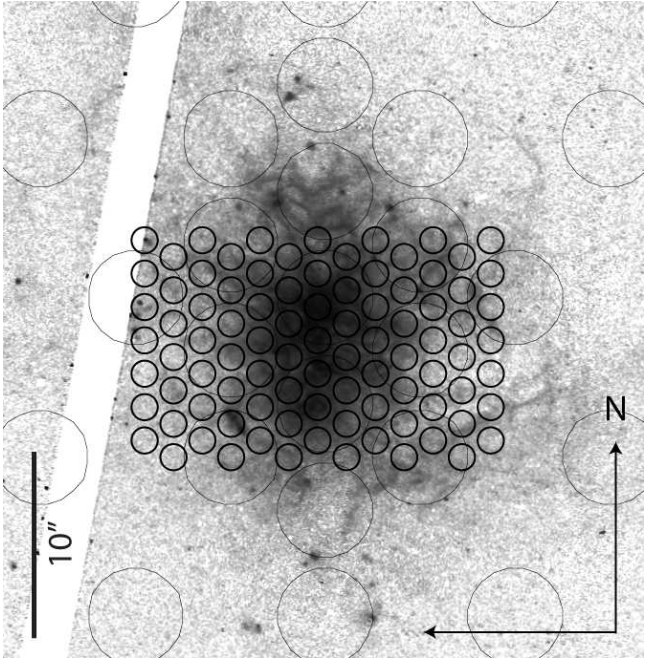}
\caption{\textit{HST}/ACS continuum subtracted H$\alpha$ image of the central regions of NGC 1140 (inverse log scaled) with DensePak footprint overlaid (bold) on the SparsePak footprint from Fig.~\ref{fig:SP_finder}.}
\label{fig:DP_finder}
\end{figure}

On 30th September 2003, we observed the nuclear regions of NGC 1140 with DensePak by centring the array on the coordinates $02^{\rm h}\,54^{\rm m}\,33\fsec43$, $-10^{\circ}\,01'\,40\farcs6$ with a PA of +90$^{\circ}$. The total exposure time was 1800~secs. Fig.~\ref{fig:DP_finder} shows the footprint of the array on the \textit{HST}/ACS F658N image. The 600~line~mm$^{-1}$ grating at an angle of 26.82$^{\circ}$ gave a wavelength range of 5260--8110~\AA\ with a dispersion of 1.40~\Apix, allowing us access to a number of optical nebular lines. Contemporaneous bias frames, flat-fields and arc calibration exposures were also observed.

\subsubsection{Reduction} \label{sect:DPreduction}
Reduction was achieved using the same methodology as outlined for SparsePak, using the {\sc ccdproc} and {\sc hydra} tasks within the {\sc noao} {\sc iraf} package. Traces of the individual spectra were identified from the master flat-field frame, and the resulting aperture identification table was used to extract the object spectra from the science frames. After flat-fielding and wavelength calibrating the dataset, we formed a representative sky spectrum by averaging the data from all four sky fibres from all three exposures. This sky spectrum was then subtracted from each of the object spectra. Final combination of the individual frames was done using {\sc imcombine} after cosmic-rays were removed with {\sc lacosmic}. The final datafiles contained 83 reduced and sky-subtracted object spectra.

In order to determine an accurate measurement of the instrumental broadening, we fitted a single Gaussian to a number of high S/N, isolated arc lines for all 83 apertures and took the average: the resulting measurement was 165\,$\pm$\,10~\kms. Note that this is considerably higher than that of the SparsePak data, and that, for the most part, the emission line profiles are unresolved.

Fig.~\ref{fig:DP_finder} shows the position of the DensePak footprint on a cut-out of the \textit{HST} F658N image with the SparsePak fibres from Fig.~\ref{fig:SP_finder}. An example reduced spectrum is shown in the bottom panel of Fig.~\ref {fig:spec_labelled}. A number of residuals remain (e.g.\ around [O\one]$\lambda$5577 and Na\one\ $\lambda\lambda$5890,5896) as a result of an imperfect sky subtraction.

Given the spectral resolution of these data, we fitted the emission lines of H$\alpha$, [N\two]$\lambda$6583 and [S\two]$\lambda\lambda$6717,6731 in each spectrum with only a single Gaussian component (\textbf{constrained to have a width greater than the aforementioned instrumental width}). Two fibres covering the brightest, central-most regions show evidence of multiple components in the H$\alpha$ emission line; in both cases they are resolved into two components separated by $\sim$100~\kms\ (these additional components are not shown in the emission line maps presented below).

\section{Emission Line Maps} \label{sect:maps}
In this section we present and describe the SparsePak and DensePak emission line maps created from the line profile fits described above. We begin by looking at the SparsePak results.

\subsection{SparsePak}\label{sect:sp_results}

\subsubsection{Kinematics}\label{sect:sp_kinematics}
\begin{figure*}
\centering
\includegraphics[width=\textwidth]{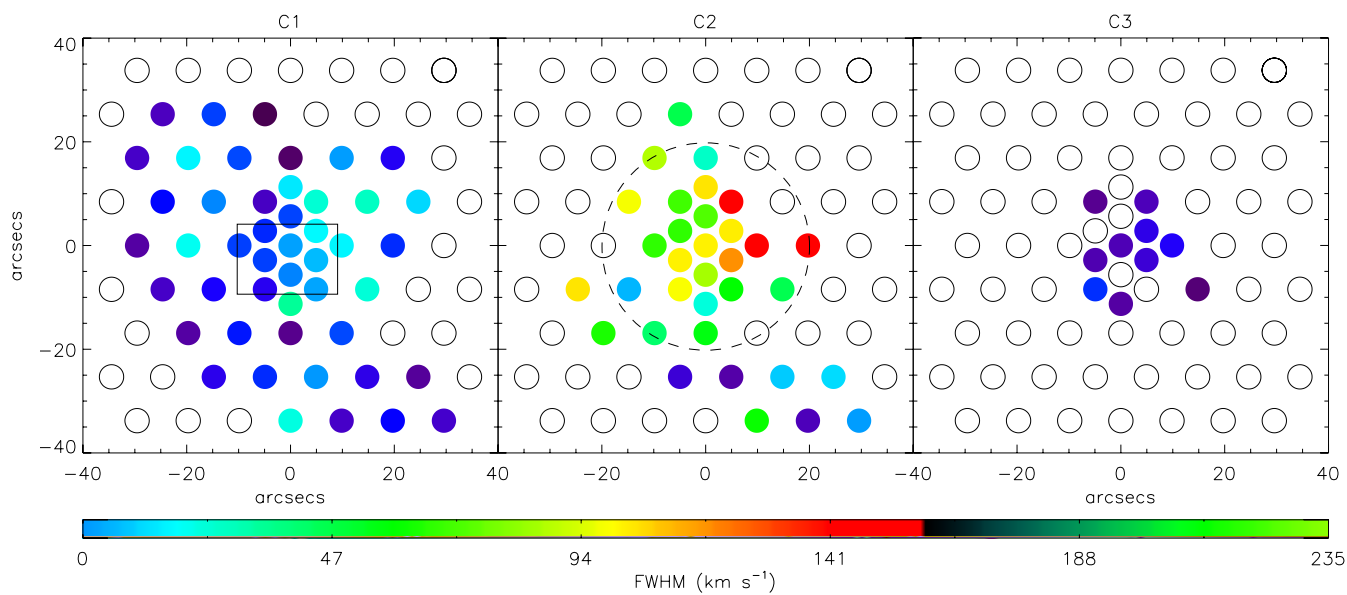}
\caption{SparsePak H$\alpha$ FWHM maps in the three H$\alpha$ line components. \emph{Left:} C1, \emph{centre:} C2 and \emph{right:} C3. A scale bar is given in units of \kms, corrected for instrumental broadening. The box in the left panel represents the DensePak footprint outline; and the dashed circle in the central panel indicates a radius of 2~kpc.}
\label{fig:sp_fwhm}
\end{figure*}

\begin{figure*}
\centering
\includegraphics[width=\textwidth]{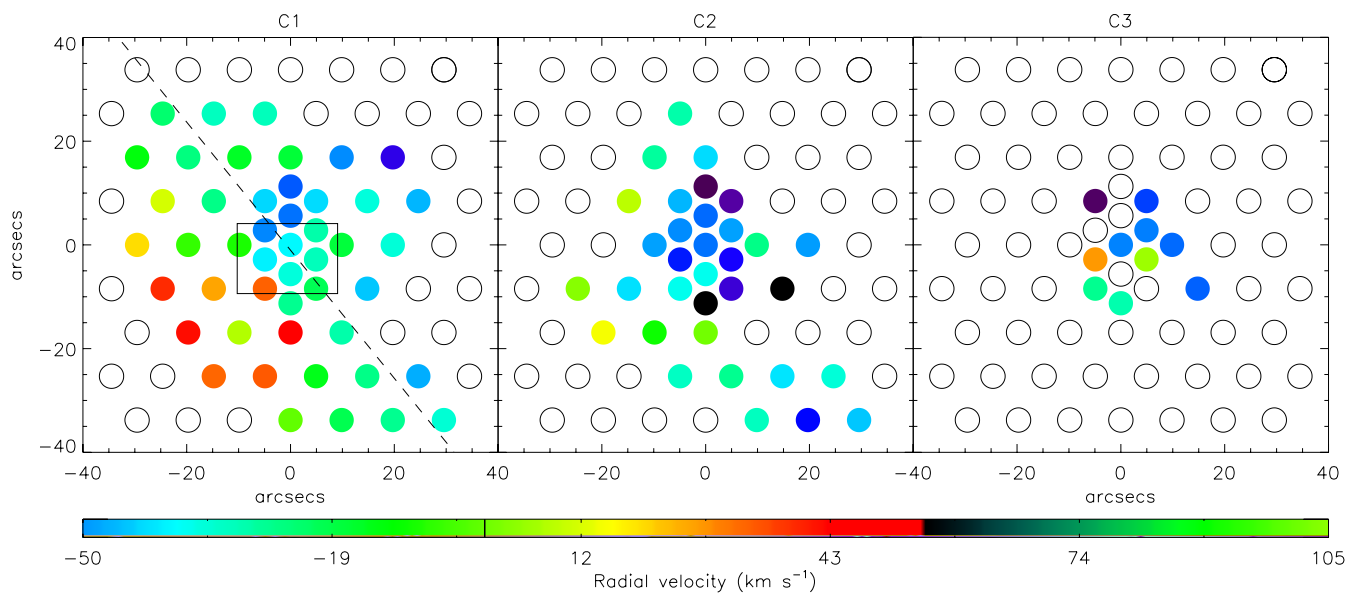}
\caption{SparsePak H$\alpha$ radial velocity maps in the three line components. \emph{Left:} C1 \citepalias[the dashed line represents the large-scale H\one\ rotation axis, \textbf{PA=$39^{\circ}$};][]{hunter94b}, \emph{centre:} C2 and \emph{right:} C3. A scale bar is given in units of \kms, relative to $v_{\rm sys}$, and the vertical bar indicates zero. The box in the left panel represents the DensePak footprint outline.}
\label{fig:sp_vel}
\end{figure*}

The spatial distribution FWHM and radial velocity of each Gaussian component identified are shown in Figs.~\ref{fig:sp_fwhm} and \ref{fig:sp_vel}, respectively. 

Narrow $\lesssim$100~\kms\ H$\alpha$ components are detected out to $>$$40''$ ($\sim$3.8~kpc) from the nucleus (Fig.~\ref{fig:sp_fwhm} left panel), corresponding well to the limit of emission seen in the deep H$\alpha$ image presented by \citetalias{hunter94b}. In C1, the imprint of the large-scale rotation of the galaxy can clearly be seen (Fig.~\ref{fig:sp_vel} left panel), and a polynomial surface fit to these data shows that the peak rotation amplitude ($\sim$20~\kms~kpc$^{-1}$) is along PA = $-51^{\circ}$ \textbf{(i.e.\ rotation axis PA = +$39^{\circ}$)}, in agreement with the H\one\ data of \citepalias{hunter94b}. In C1, the inner $\sim$$20\times 20''$ of the galaxy shows hints of rotating at a different position angle. This is discussed further in light of the DensePak data presented in Section~\ref{sect:dp_kinematics}. We would also like to note that the movement of H$\alpha$-emitting gas associated with the chain of H\two\ regions (Fig~\ref{fig:SP_finder}) is consistent with the rest of the halo \citepalias[and with the findings of][]{hunter94b}.

An underlying broad component (C2; 100--230~\kms) can be identified throughout the disk and inner halo, extending out to a radius of $\gtrsim$2~kpc (Fig.~\ref{fig:sp_fwhm} central panel). In general, broader line widths are seen on the western side of the galaxy. This is mirrored in both C1 and C2, where C2 widths increase up to FWHM = 230~\kms\ (Fig.~\ref{fig:sp_fwhm}; see also Fig.~\ref{fig:sp_eg_fits} top-right panel). No broad H$\alpha$ emission is associated with the chain of H\two\ regions towards the south-west. Here only two narrow components are present (note that the second, fainter component has been assigned to C2 -- our assignment convention is described in Section~\ref{sect:line_fitting}). In general, the radial velocities of the broad C2 component are blueshifted by $\sim$10--50~\kms\ compared to C1, suggesting that it is associated with gas expanding out of the star forming zones. These speeds are more typical for expanding supershells as compared to the much larger velocities seen in galactic winds.

At this point we would like to note that in NGC 1569 we compared the range of emission line radial velocities measured in individual spaxels over one $5\times 3.5$~arcsecs GMOS IFU field-of-view \citep{westm07b} with the width of lines from SparsePak fibres that covered approximately the same area \citep{westm08}, and found that the SparsePak FWHM measurements were very likely artificially elevated at the level of a few tens of \kms\ by multiple (unresolved) emission components along the line-of-sight. An effect similar to this is also very likely to be at work here, particularly in the central regions. Here, the presence of complex filamentary emission line structures implies multiple overlapping shells or dynamical components on scales less than the DensePak or SparsePak fibre sizes.

The presence of a second narrow component (assigned to C3 where a broad component is identified and C2 otherwise) is indicative of expanding gas motions. These split lines are seen within the nuclear regions (central part of array) and in the chain of H\two\ regions, where the velocity differences between the two components are 10--60~\kms. These kind of velocity differences are consistent with those expected from regions of intense star formation, and suggest the presence of outflowing gas.

\subsubsection{Nebular diagnostics}\label{sect:sp_diagnostics}
The [S\two]$\lambda\lambda 6717,6731$ lines were used to derive the electron densities of the ionized gas. For the majority of the galaxy where the the [S\two] lines were detected (the central $\sim$2~kpc and along the chain of H\two\ regions), the derived values fell below the low-density limit at $\sim$100~\cmt. However, in an area encompassing the central few spaxels and those extending $\sim$$20''$ to the south, we find C1 densities rise to a few 100~\cmt. In the spaxels located in the central part of the array (inner $\sim$1~kpc) a second [S\two] component could also be identified. The densities measured in this component were again mostly at or just above the low-density limit.

These results are consistent with those of previous studies: \citetalias{hunter94b} find densities of a few 100~\cmt\ in the central 20--40$''$ of the galaxy, and \citet{moll07} use oxygen line diagnostics to find a density of $60\pm 50$~\cmt\ in the nebular region directly surrounding starburst knot A (the northern star cluster complex). \citet{burgh06} find electron densities of just over 100~\cmt\ from the [S\two] ratio measured in the nuclear regions.

\begin{figure*}
\centering
\includegraphics[width=\textwidth]{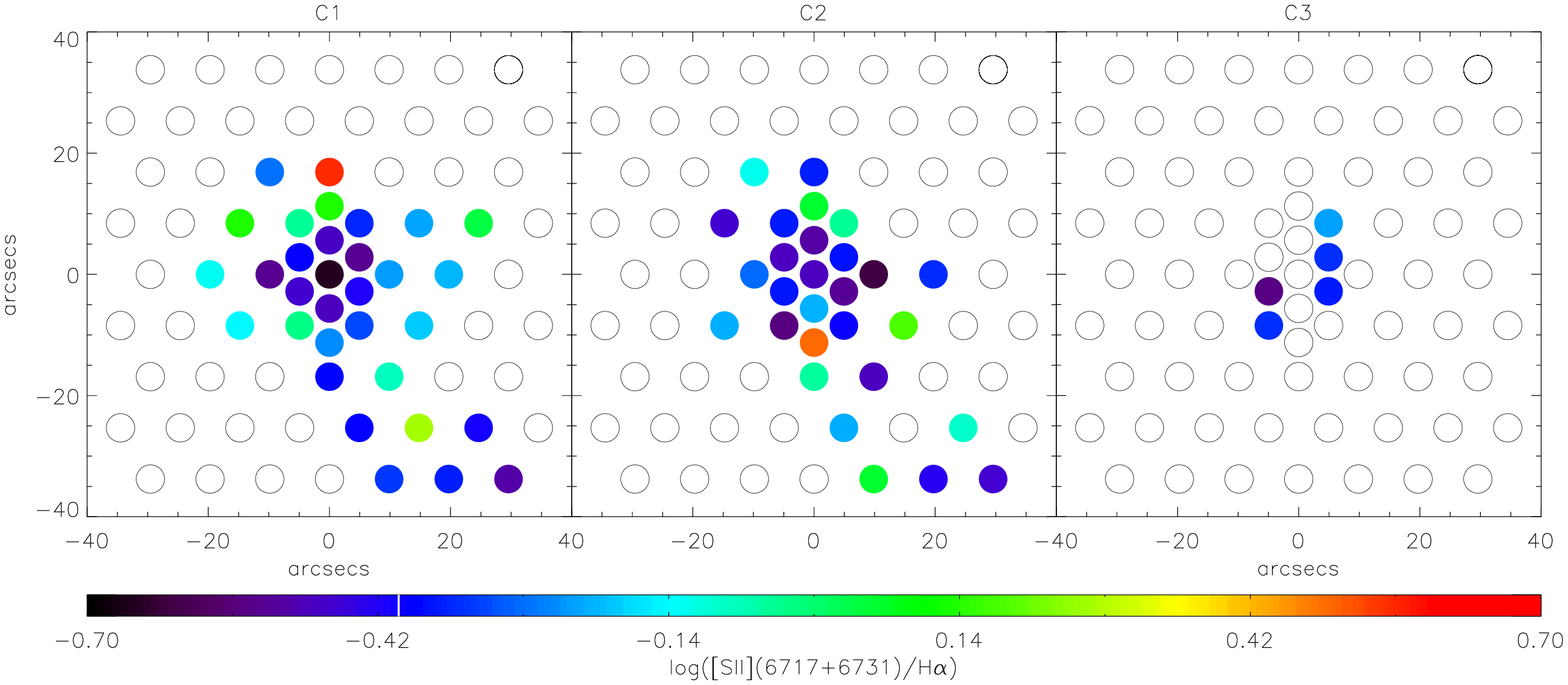}
\caption{SparsePak [S\two]$\lambda\lambda 6717,6731$/H$\alpha$ ratio maps in the three line components. \emph{Left:} C1, \emph{centre:} C2 and \emph{right:} C3. The vertical line in the scale bar represents a fiducial ratio (log([S\two]/H$\alpha$) = $-0.4$) above which it is expected that a significant proportion of the ionization is achieved through non-photoionizing processes.}
\label{fig:sp_SII_Ha}
\end{figure*}

The forbidden/recomination line flux ratio of [S\two]($\lambda$6717+$\lambda$6731)/H$\alpha$ can be used as an indicator of the number of ionizations per unit volume, and thus the ionization parameter, $U$ \citep{veilleux87,dopita00,dopita06b}. [S\two]/H$\alpha$ is also particularly sensitive to shock ionization because relatively high-density, partially ionized regions form behind shock fronts which emit strongly in [S\two] thus producing an enhancement in [S\two]/H$\alpha$ \citep{dopita97, oey00, osterbrock06}. Maps of this ratio in the three line components are shown in Fig.~\ref{fig:sp_SII_Ha}. In the scale bar, we have indicated a fiducial value of log([S\two]/H$\alpha$) = $-0.4$ above which non-photoionized emission is thought to play a dominant role \citep{dopita95, kewley01, calzetti04}. In environments such as these, the most likely excitation mechanism after photoionization is that of shocks. Outside of the central 2~kpc region, the ratios in all three line components rise significantly above this non-photoionization threshold. This suggests that shocks dominate the gas excitation in the halo (as previously noted by \citealt{calzetti04} in the outer regions of other moderate luminosity dwarf starburst galaxies) and thus may play a central role in driving the gas outward.

\begin{figure*}
\centering
\includegraphics[width=\textwidth]{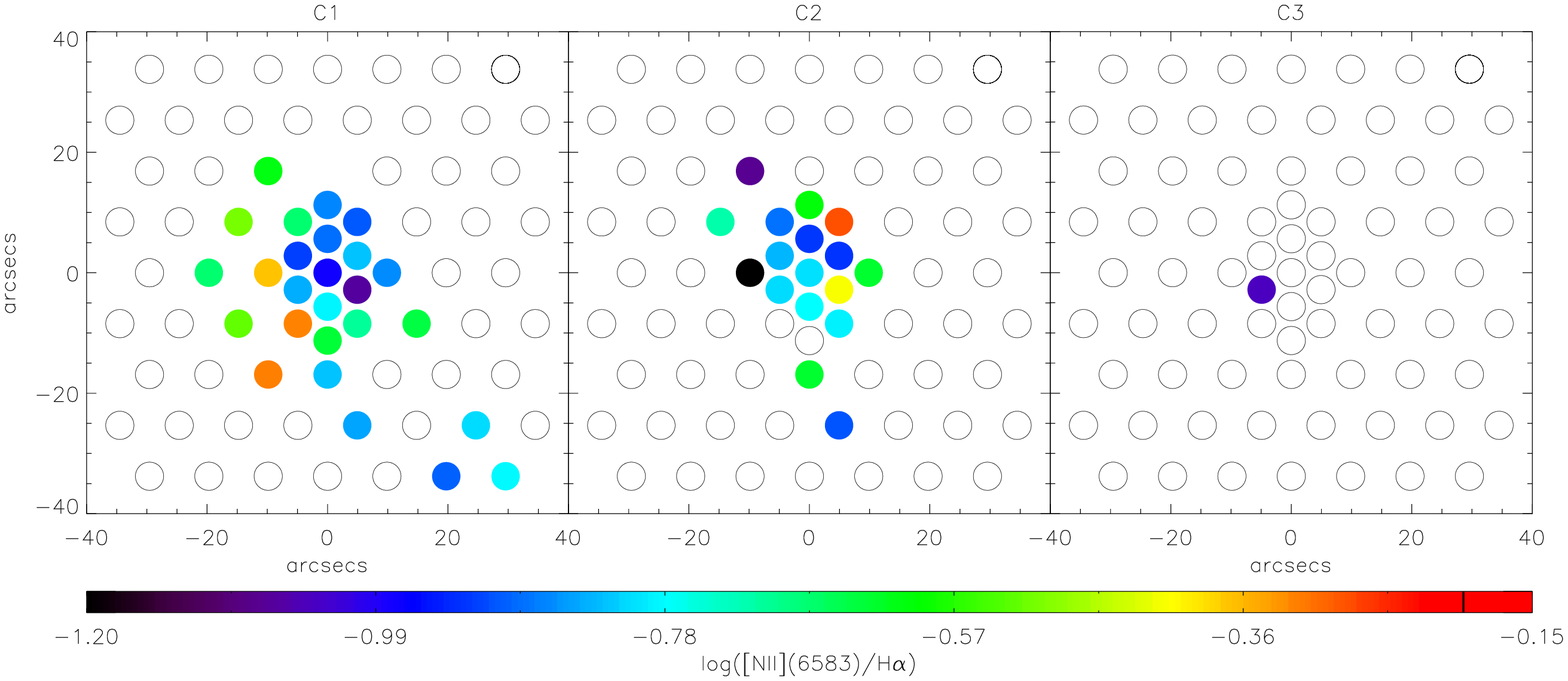}
\caption{SparsePak [N\two]$\lambda 6583$/H$\alpha$ ratio maps in the three line components. \emph{Left:} C1, \emph{centre:} C2 and \emph{right:} C3. The vertical line in the scale bar represents a fiducial ratio (log([N\two]/H$\alpha$) = $-0.2$) above which it is expected that a significant proportion of the ionization is achieved through non-photoionizing processes.}
\label{fig:sp_NII_Ha}
\end{figure*}

The ratio of [N\two]$\lambda$6583/H$\alpha$ can be used in a similar fashion to [S\two]/H$\alpha$ as a diagnostic of the gas excitation level and to search for the presence of shocks \citep{veilleux87, dopita95}. Fig.~\ref{fig:sp_NII_Ha} shows maps of this ratio for all three components; here the marker in the scale bar represents a fiducial non-photoionized threshold of log([N\two]/H$\alpha$) = $-0.1$. Again we see that the line ratios increase from the central region outwards, but in this case never rise above our adopted non-photoionization threshold. Here we should note that metallicity plays a strong part in setting the H\two\ region/shock boundary for these diagnostic ratios, in the sense that a lower N abundance (as would be expected at the low metallicity of NGC 1140) would decrease the non-photoionization threshold \citep{dopita06b}.

Again our results agree very well with previous studies: \citetalias{hunter94b} measured log([S\two]/H$\alpha$) $\sim$ $-0.7$ to $-0.3$ and log([N\two]/H$\alpha$) $\sim$ $-1$ to $-0.7$ in the central regions of NGC 1140, and \citet{burgh06} found [S\two]/H$\alpha$ to vary with radius from a minimum of log([S\two]/H$\alpha$) = $-0.9$ at the peak of the H$\alpha$ emission, to log([S\two]/H$\alpha$) = $-0.3$ at the edges of the galaxy.

\subsection{DensePak results}\label{sect:dp_results}
As described in Section~\ref{sect:DPreduction}, the emission lines in the DensePak spectra were almost all unresolved meaning that a single Gaussian component fit was all that was required. The spatial distribution of H$\alpha$ flux and radial velocity, and the [S\two]/H$\alpha$ and [N\two]/H$\alpha$ line ratios, resulting from these fits are shown in Fig.~\ref{fig:dp_maps}. The location of the DensePak array on the \textit{HST} H$\alpha$ image is shown in Fig.~\ref{fig:DP_finder}.

\subsubsection{Kinematics}\label{sect:dp_kinematics}
Fig.~\ref{fig:dp_maps}b shows the H$\alpha$ radial velocity map. It is clear that the velocity gradient observed does not follow what is seen on large scales in, for example, our SparsePak maps (Fig.~\ref{fig:sp_vel}). It is instead offset by approximately 90$^{\circ}$ to a PA of $\sim$+40$^{\circ}$. This corroborates what was hinted at in the SparsePak data (Section~\ref{sect:sp_kinematics}; and in the long-slit data of \citealt{burgh06}), and suggests the presence of a counter-rotating core within the central $\sim$$20\times 20''$ with a rotation axis aligned roughly perpendicular to the main body rotation axis. This was also found in the long-slit data of \citet{burgh06}.

\subsubsection{Nebular diagnostics}\label{sect:dp_diagnostics}
Most DensePak spaxels exhibit [S\two]-derived electron densities below the low-density limit (100~\cmt), although a small number show densities of a few hundred \cmt. These results are consistent with our SparsePak measurements and with previous studies as described above.

\begin{figure*}
\centering
\begin{minipage}{8cm}
\includegraphics[width=8cm]{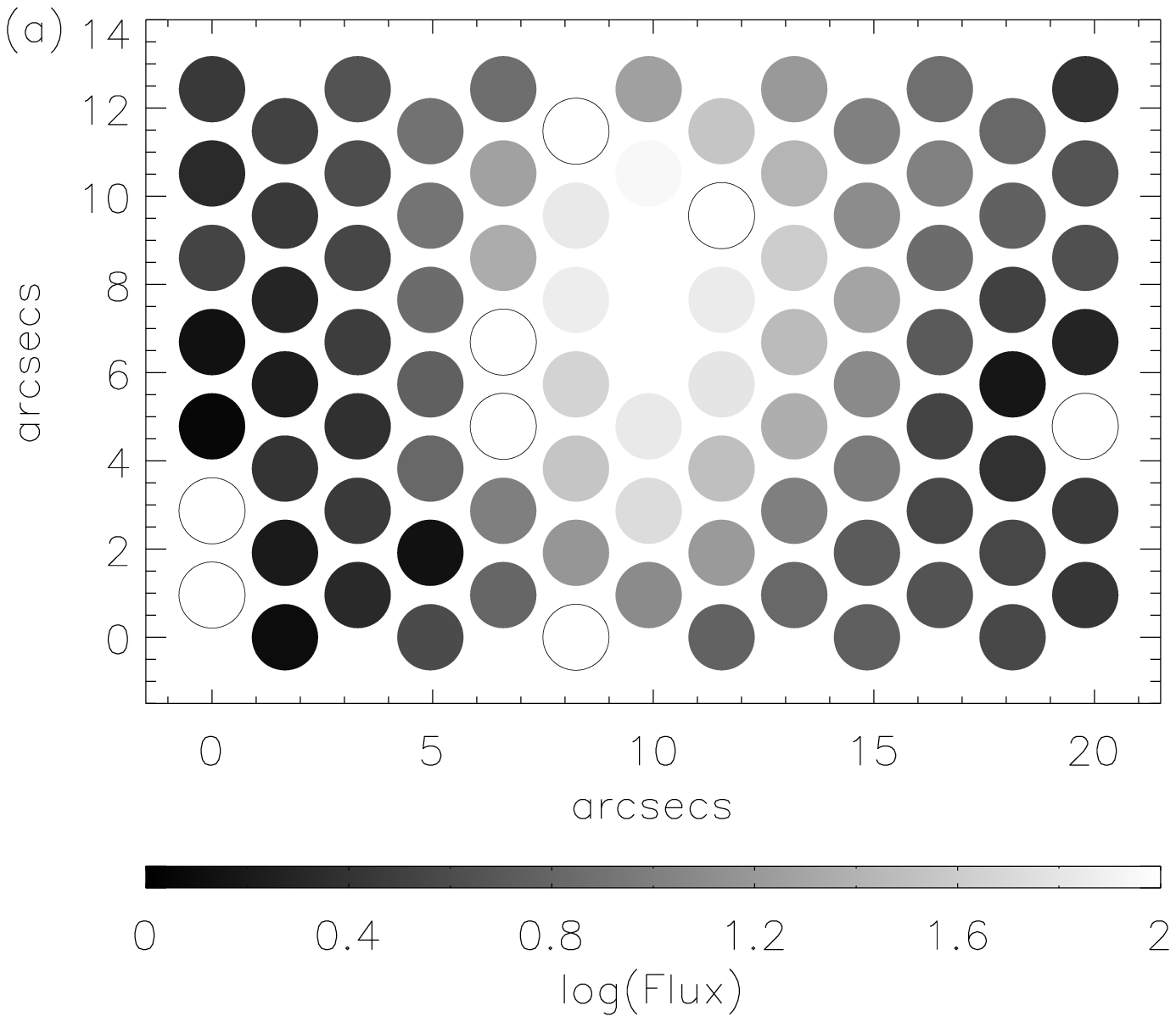}
\end{minipage}
\hspace{0.2cm}
\begin{minipage}{8cm}
\includegraphics[width=8cm]{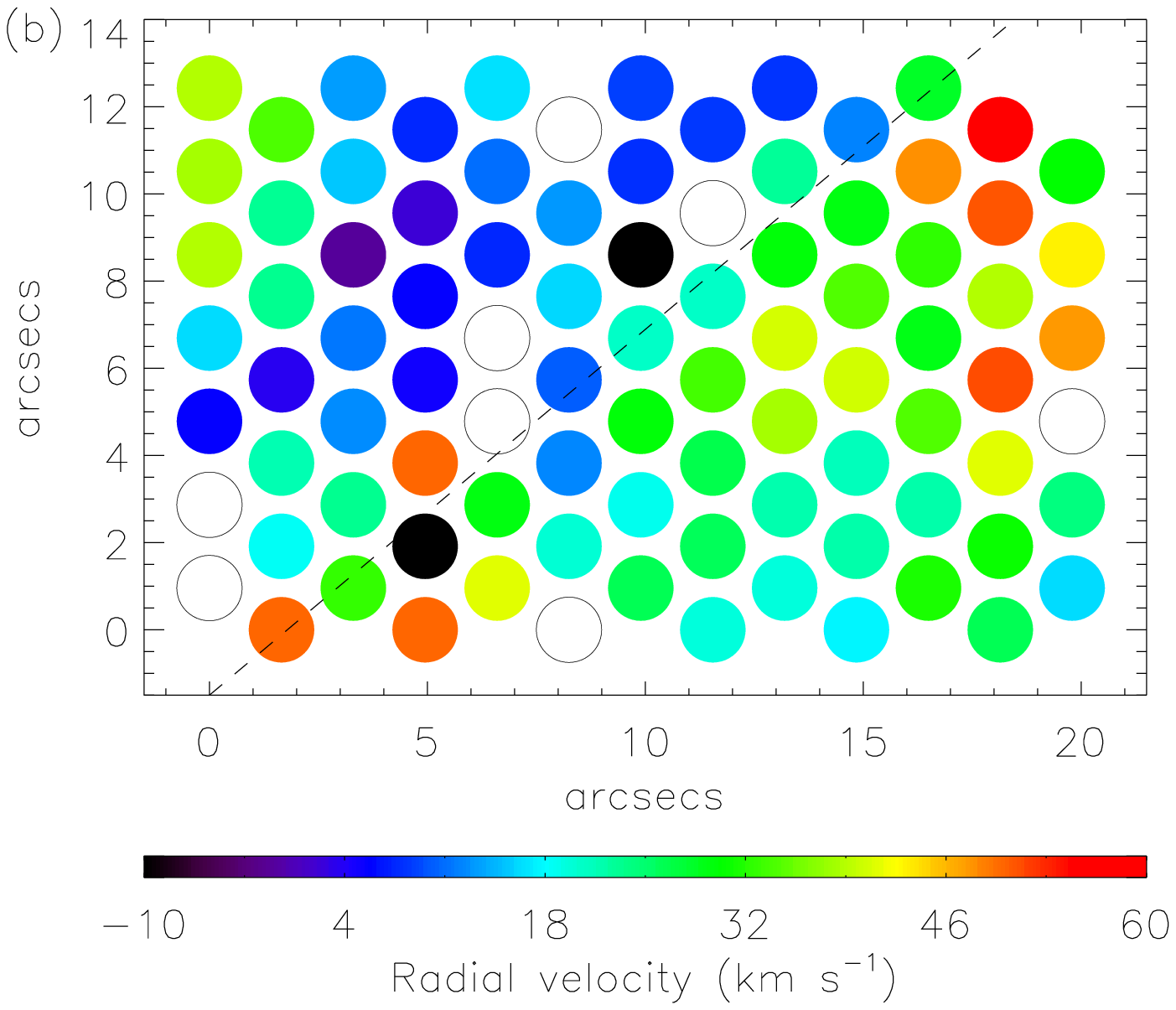}
\end{minipage}
\vspace{0.0cm}
\begin{minipage}{8cm}
\includegraphics[width=8cm]{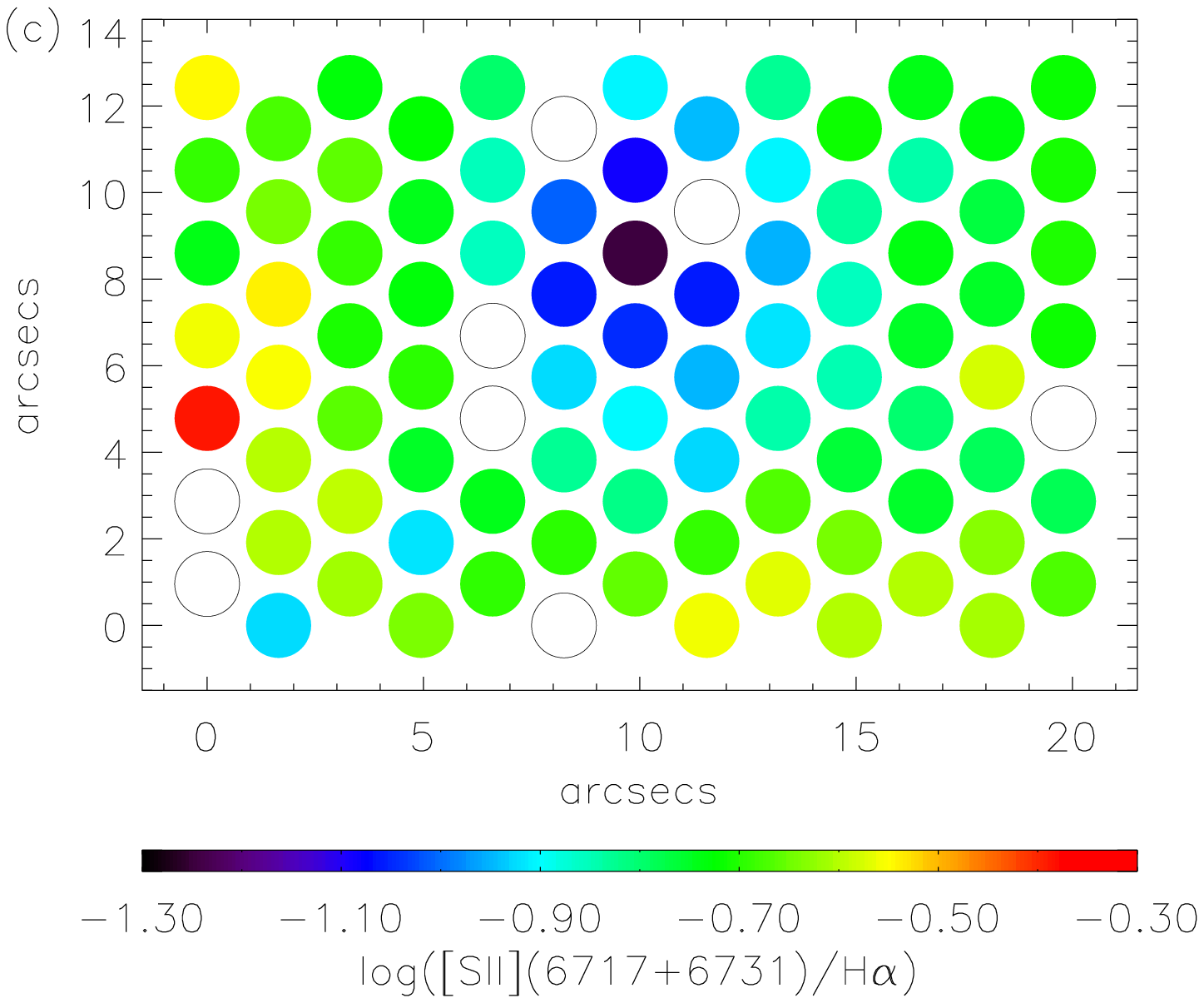}
\end{minipage}
\hspace{0.2cm}
\begin{minipage}{8cm}
\includegraphics[width=8cm]{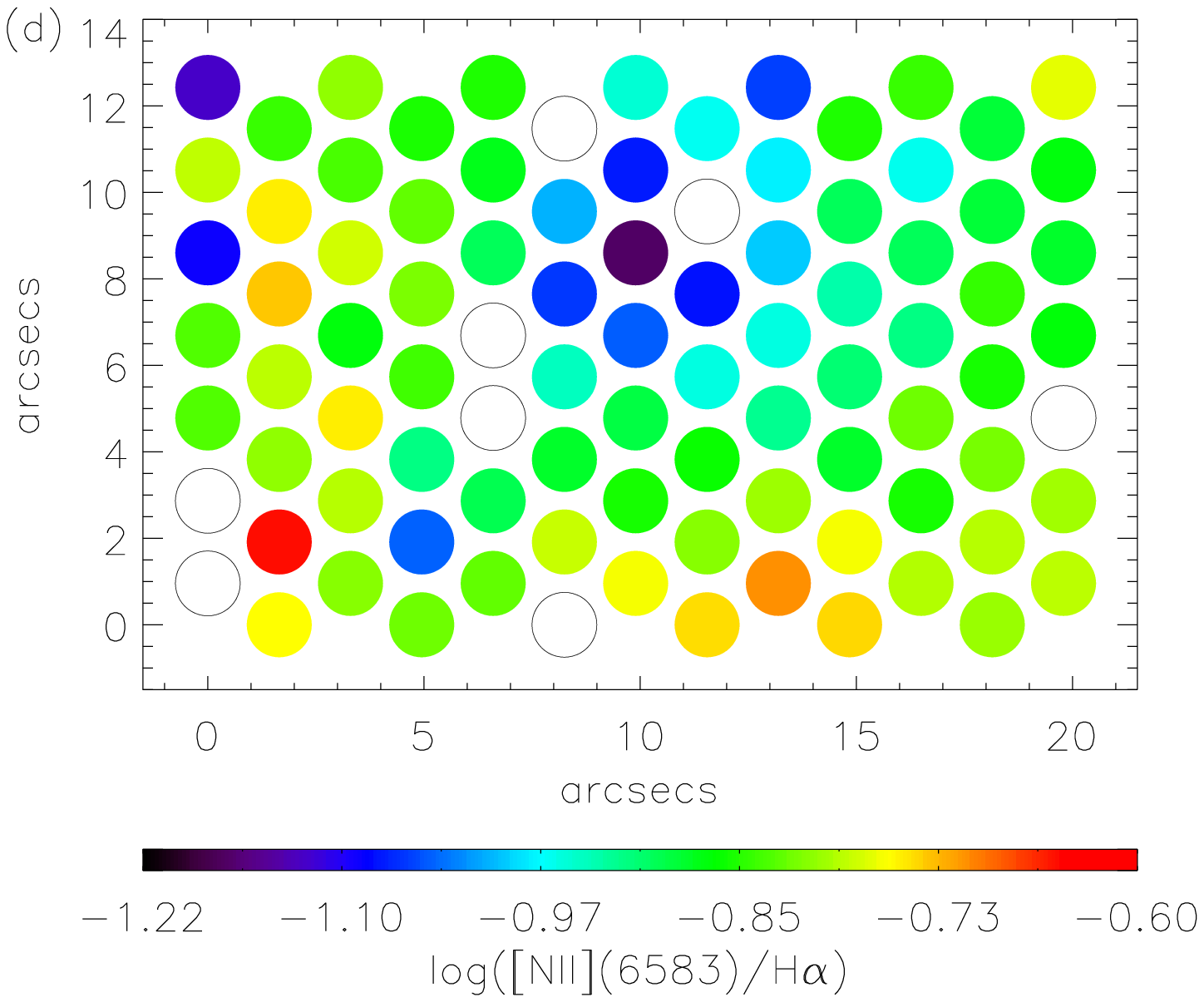}
\end{minipage}
\caption{DensePak maps: (a) H$\alpha$ fluxes; (b) H$\alpha$ radial velocities (the dashed line indicates the axis of rotation of the counter-rotating core, \textbf{PA $\approx$ $-50^{\circ}$}); (c) [S\two]/H$\alpha$ ratio; (d) [N\two]/H$\alpha$ ratio. Scale bars are shown for each map individually. Unlike with SparsePak, only one line component can be detected with DensePak due to the much lower spectral resolution of these observations.}
\label{fig:dp_maps}
\end{figure*}

In both [S\two]/H$\alpha$ and [N\two]/H$\alpha$ (Fig.~\ref{fig:dp_maps}c and d) the lowest ratios are coincident with the location of the H$\alpha$ peak (Fig.~\ref{fig:dp_maps}a), then increase radially outward (the similarity of the two maps is striking). No evidence for shocked line ratios (in either [S\two]/H$\alpha$ or [N\two]/H$\alpha$) are found in these central regions. Our measurements are again very consistent with those from the SparsePak data (once the finer spatial sampling has been taken into account), and with previous studies (\citetalias{hunter94b}; \citealt{burgh06}).

\section{Discussion and conclusions} \label{sect:disc}

In this paper we have presented deep spatially-resolved optical spectra of NGC 1140 obtained with the WIYN SparsePak and DensePak instruments. The different spatial resolutions and coverage of the two sets of observations have allowed us to investigate the properties and kinematics of the warm ionized gas within both the central regions of the galaxy and the inner halo, and have revealed a number of interesting results. We will now summarise these and discuss their implications in the context of the galaxy as a whole.

\subsection{Kinematical structure of the disk and halo}
We have measured the position angle of the H$\alpha$ rotation axis for main body of the galaxy, and find it to be consistent with the H\one\ rotation axis at PA = $39^{\circ}$ \citepalias{hunter94b}.

The peak of the optical emission lies roughly at the centre of the integrated H\one\ morphology \citepalias{hunter94b}. The H\one\ ridge feature is oriented roughly perpendicularly to the major axis, and contains the peak of the H\one\ emission, which lies to the south-west of the optical peak, coincident with the chain of H\two\ regions \citepalias{hunter94b}. Despite their appearance, the chain of H\two\ regions located towards the south-west are not part of a tidal tail (which might be expected to be kinematically distinct), but appear to rotate together with the rest of the galaxy. This makes sense since we know they are coincident with the H\one\ ridge identified by \citetalias{hunter94b}, and the peak of the H\one\ emission. What do the chain of H\two\ regions represent? Split H$\alpha$ lines with velocity separations of a few tens of \kms\ are seen at the location of the chain of H\two\ regions indicating that they are young, star forming, and dynamically active. It is possible that we are witnessing a sequential progression of star formation through the disk (perhaps the H\two\ regions represent the stub of a tidally induced spiral arm), and that in a few Myr the galaxy will look elongated in H$\alpha$ along the $+22^{\circ}$ ridge direction as star formation takes hold.

Our data clearly show that the ionized gas in the central $20\times 20$~arcsecs ($2\times 2$~kpc) region of the galaxy is kinematically decoupled from the rest of the system, and rotates at a PA approximately perpendicular to that of the main body of the galaxy at $+40^{\circ}$. This finding of yet another misaligned structural component is not unsurprising, and only adds weight to the interaction/merger scenario proposed by \citetalias{hunter94b}.

\subsection{The starburst environment and the properties of the outflowing gas}
We see no evidence of coherent large-scale outflows within NGC 1140. Instead, the H$\alpha$ morphology (Fig~\ref{fig:SP_finder}), the fact that we find multiple narrow line components with velocity separations of 10--60~\kms\ within a radius of $\sim$1--1.5~kpc, and the fact that beyond $\sim$2~kpc the [S\two]/H$\alpha$ ratios rise significantly implying a strong contribution to the excitation levels by shocks, all suggest that locally intense star formation is driving material outwards in the form of a series of interacting superbubbles/shells (i.e.\ the ``froth''). The lack of any coherent wind may be due to the disturbed structure of the galaxy disk, and/or the modest thermal pressures in the starburst core (implied by the low electron densities) and/or the low global average star formation rate ($<$1~\Msol~yr$^{-1}$; \citetalias{hunter94b}; \citealt{moll07}).



A broad component (100\,$\lesssim$\,FWHM\,$\lesssim$\,230~\kms) to the H$\alpha$ line is seen throughout the galaxy disk out to $>$2~kpc. Broad emission line components have been observed before in the H$\beta$ line profiles extracted from the nebula regions surrounding the SSC knots A and B by \citet{moll07}, and in many other nearby starburst galaxies (e.g.~\citealt{izotov96, homeier99, marlowe95, mendez97}; \citealt*{sidoli06}). For many years the nature of the energy source for these broad lines has been contested, but recent detailed IFU studies of the ionized ISM in the starburst galaxies NGC 1569 and M82 \citep{westm07a, westm07b, westm09a, westm09b} have allowed us to address this problem. Accurately decomposing the emission line profiles and mapping out their properties led us to the conclusion that the broad underlying component is produced in turbulent mixing layers \citep[TMLs;][]{slavin93, binette09} on the surfaces of cool gas knots, set up by the impact of the ionizing radiation and fast-flowing winds from the YMCs \citep{pittard05}. In both these galaxies, we found evidence for a highly fragmented ISM that provides copious cloud surfaces on which TMLs can form thus explaining the pervasiveness of this broad component. The existence of multiple YMCs in the central region, the complex, ``frothy'' H$\alpha$ morphology of the disk and halo (Fig.~\ref{fig:SP_finder}), and the similarity in width and morphology of the broad component observed here, strongly suggest a similar origin for this component. The same conclusion was reached by \citet{moll07}.

Can we determine if the radius at which we stop seeing the broad component is physical or simply an effect of the data? Examination of a number of H$\alpha$ line profiles with similar S/N levels from spaxels near this boundary reveals slight evidence for the former: spaxels 82 (coordinates +20, 0) and 37 ($-25$, $-8$) both exhibit a broad H$\alpha$ component and have integrated H$\alpha$ S/N ratios of $\sim$55 and $\sim$37, respectively, whereas spaxels 28 ($-20$, 0) and 63 (15,8), with integrated H$\alpha$ S/N ratios of $\sim$85 and $\sim$45, respectively, do not. Finding a physical limit to the broad component region echoes what we found in NGC 1569 \citep{westm08}. Here we found that the radius at which the broad line component ceased to exist roughly corresponded to the point that the H$\alpha$ profile started showing a secondary narrow component. Clearly the situation is somewhat different in NGC 1140 since the broad component region is much larger ($\sim$2~kpc vs.\ 500~pc), and we see split narrow components within the central regions, but this limit to the broad component region may still mark a significant transition point in the development of the galactic outflow, where turbulent motion becomes less dominant. Further investigation at higher spatial resolution and S/N and with comparison to deeper H$\alpha$ imaging would be needed to better quantify this potential transition region.

We can also try to assess the relationship between the state of the outflow in NGC 1140 and that of NGC 1569 and the well-known starburst M82. On galaxy size-scales, the NGC 1569 outflow exhibits an irregular supershell morphology with very little collimation or preferred outflow direction \citep{hunter93, martin98, westm08}. M82's outflow, on the other hand, consists of two large bi-polar plumes of outflowing gas that are relatively well structured and collimated \citep{shopbell98, ohyama02, westm09a}. The halo of NGC 1140, with its complex, frothy morphology, seems mid-way between these two examples.

Firstly, NGC 1140 is almost three orders of magnitude more massive than NGC 1569 and has a rotational velocity $v_{\rm rot}\sim 100$~\kms\ \citepalias{hunter94b} compared to $\sim$35~\kms\ in NGC 1569 \citep{stil02}. Since the escape speed, $v_{\rm esc}$, scales as $v^2 R$, $v_{\rm esc}$ in NGC 1140 is $>$3 times greater than that in NGC 1569. The two galaxies have similar star formation rates per unit area so it may not be surprising that the burst in NGC 1140 is producing less in the way of an outflow. 

NGC 1140 and M82 have similar $v_{\rm rot}$ and sizes, so should have roughly similar values of $v_{\rm esc}$. However, both the thermal and turbulent pressures at the base of the M82 wind are $>$10 times that in NGC 1140 meaning that M82 must have a significantly higher energy density in its ISM and is therefore able to drive a large-scale superwind in a way NGC 1140 cannot. The presence of a broad emission line component in both cases suggests that despite this difference, the underlying structure of cloud boundaries and turbulent mixing layers are similar, and that these features are not directly linked to the magnitude of the winds.

\section*{Acknowledgments}
We thank the anonymous referee for comments that led to an improvement of the clarity of this paper. 
MSW would like to thank Daniel Harbeck and the staff at the WIYN Observatory for their support during the observing runs, and Phillip Cigan for reducing the DensePak observations.
MSW also thanks the University of Wisconsin--Madison for the warm hospitality received in support of this project. JSG's research was partially funded by the National Science Foundation through grant AST-0708967 to the University of Wisconsin-Madison.

\bibliographystyle{mn2e}
\bibliography{/Users/msw/Documents/work/references}

\begin{thebibliography}{}

\bibitem[\protect\citeauthoryear{{Aguirre}, {Hernquist}, {Schaye}, {Weinberg},
  {Katz} \& {Gardner}}{{Aguirre} et~al.}{2001}]{aguirre01a}
{Aguirre} A.,  {Hernquist} L.,  {Schaye} J.,  {Weinberg} D.~H.,  {Katz} N.,
  {Gardner} J.,  2001, \apj, 560, 599

\bibitem[\protect\citeauthoryear{{Barden}, {Sawyer} \& {Honeycutt}}{{Barden}
  et~al.}{1998}]{barden98}
{Barden} S.~C.,  {Sawyer} D.~G.,    {Honeycutt} R.~K.,  1998, in {D'Odorico}
  S.,  ed., Proc. SPIE Vol. 3355, p. 892-899, Optical Astronomical
  Instrumentation, {Integral field spectroscopy on the WIYN telescope using a
  fiber array}.
pp 892--899

\bibitem[\protect\citeauthoryear{{Bershady}, {Andersen}, {Harker}, {Ramsey} \&
  {Verheijen}}{{Bershady} et~al.}{2004}]{bershady04}
{Bershady} M.~A.,  {Andersen} D.~R.,  {Harker} J.,  {Ramsey} L.~W.,
  {Verheijen} M.~A.~W.,  2004, \pasp, 116, 565

\bibitem[\protect\citeauthoryear{{Binette}, {Drissen}, {Ubeda}, {Raga},
  {Robert} \& {Krongold}}{{Binette} et~al.}{2009}]{binette09}
{Binette} L.,  {Drissen} L.,  {Ubeda} L.,  {Raga} A.~C.,  {Robert} C.,
  {Krongold} Y.,  2009, arXiv 0902.3689

\bibitem[\protect\citeauthoryear{{Burgh}, {Gallagher} III, {Nordsieck},
  {Percival}, {Smith}, {O'Donoghue}, {Buckley} \& {Loaring}}{{Burgh}
  et~al.}{2006}]{burgh06}
{Burgh} E.~B.,  {Gallagher} III J.~S.,  {Nordsieck} K.~H.,  {Percival} J.~W.,
  {Smith} M.~P.,  {O'Donoghue} D.,  {Buckley} D.~A.,    {Loaring} N.~S.,  2006,
  in Bulletin of the American Astronomical Society Vol.~38, {SALT/RSS Longslit
  Spectroscopy of the NGC 1140 Starburst}.
p.~93

\bibitem[\protect\citeauthoryear{{Calzetti}, {Harris}, {Gallagher}, {Smith},
  {Conselice}, {Homeier} \& {Kewley}}{{Calzetti} et~al.}{2004}]{calzetti04}
{Calzetti} D.,  {Harris} J.,  {Gallagher} J.~S.,  {Smith} D.~A.,  {Conselice}
  C.~J.,  {Homeier} N.,    {Kewley} L.,  2004, \aj, 127, 1405

\bibitem[\protect\citeauthoryear{{Cen}, {Nagamine} \& {Ostriker}}{{Cen}
  et~al.}{2005}]{cen05}
{Cen} R.,  {Nagamine} K.,    {Ostriker} J.~P.,  2005, \apj, 635, 86

\bibitem[\protect\citeauthoryear{{Cid Fernandes}, {Le{\~a}o} \& {Lacerda}}{{Cid
  Fernandes} et~al.}{2003}]{cidfernandes03}
{Cid Fernandes} R.,  {Le{\~a}o} J.~R.~S.,    {Lacerda} R.~R.,  2003, \mnras,
  340, 29

\bibitem[\protect\citeauthoryear{{de Grijs}, {Smith}, {Bunker}, {Sharp},
  {Gallagher}, {Anders}, {Lan{\c c}on}, {O'Connell} \& {Parry}}{{de Grijs}
  et~al.}{2004}]{degrijs04}
{de Grijs} R.,  {Smith} L.~J.,  {Bunker} A.,  {Sharp} R.~G.,  {Gallagher}
  J.~S.,  {Anders} P.,  {Lan{\c c}on} A.,  {O'Connell} R.~W.,    {Parry} I.~R.,
   2004, \mnras, 352, 263

\bibitem[\protect\citeauthoryear{{De Young} \& {Heckman}}{{De Young} \&
  {Heckman}}{1994}]{deyoung94}
{De Young} D.~S.,  {Heckman} T.~M.,  1994, \apj, 431, 598

\bibitem[\protect\citeauthoryear{{Dekel} \& {Silk}}{{Dekel} \&
  {Silk}}{1986}]{dekel86}
{Dekel} A.,  {Silk} J.,  1986, \apj, 303, 39

\bibitem[\protect\citeauthoryear{{Dimeo}}{{Dimeo}}{2005}]{dimeo}
{Dimeo} R.,  2005, PAN User Guide

\bibitem[\protect\citeauthoryear{{Dopita}}{{Dopita}}{1997}]{dopita97}
{Dopita} M.~A.,  1997, \apjl, 485, L41

\bibitem[\protect\citeauthoryear{{Dopita}, {Fischera}, {Sutherland}, {Kewley},
  {Leitherer}, {Tuffs}, {Popescu}, {van Breugel} \& {Groves}}{{Dopita}
  et~al.}{2006}]{dopita06b}
{Dopita} M.~A.,  {Fischera} J.,  {Sutherland} R.~S.,  {Kewley} L.~J.,
  {Leitherer} C.,  {Tuffs} R.~J.,  {Popescu} C.~C.,  {van Breugel} W.,
  {Groves} B.~A.,  2006, \apjs, 167, 177

\bibitem[\protect\citeauthoryear{{Dopita}, {Kewley}, {Heisler} \&
  {Sutherland}}{{Dopita} et~al.}{2000}]{dopita00}
{Dopita} M.~A.,  {Kewley} L.~J.,  {Heisler} C.~A.,    {Sutherland} R.~S.,
  2000, \apj, 542, 224

\bibitem[\protect\citeauthoryear{{Dopita} \& {Sutherland}}{{Dopita} \&
  {Sutherland}}{1995}]{dopita95}
{Dopita} M.~A.,  {Sutherland} R.~S.,  1995, \apj, 455, 468

\bibitem[\protect\citeauthoryear{{Homeier} \& {Gallagher}}{{Homeier} \&
  {Gallagher}}{1999}]{homeier99}
{Homeier} N.~L.,  {Gallagher} J.~S.,  1999, \apj, 522, 199

\bibitem[\protect\citeauthoryear{{Hunter}, {Hawley} \& {Gallagher}}{{Hunter}
  et~al.}{1993}]{hunter93}
{Hunter} D.~A.,  {Hawley} W.~N.,    {Gallagher} J.~S.,  1993, \aj, 106, 1797

\bibitem[\protect\citeauthoryear{{Hunter}, {O'Connell} \& {Gallagher}}{{Hunter}
  et~al.}{1994}]{hunter94a}
{Hunter} D.~A.,  {O'Connell} R.~W.,    {Gallagher} J.~S.,  1994, \aj, 108, 84

\bibitem[\protect\citeauthoryear{{Hunter}, {van Woerden} \& {Gallagher}
  III}{{Hunter} et~al.}{1994}]{hunter94b}
{Hunter} D.~A.,  {van Woerden} H.,    {Gallagher} III J.~S.,  1994, \apjs, 91,
  79

\bibitem[\protect\citeauthoryear{{Izotov}, {Dyak}, {Chaffee}, {Foltz},
  {Kniazev} \& {Lipovetsky}}{{Izotov} et~al.}{1996}]{izotov96}
{Izotov} Y.~I.,  {Dyak} A.~B.,  {Chaffee} F.~H.,  {Foltz} C.~B.,  {Kniazev}
  A.~Y.,    {Lipovetsky} V.~A.,  1996, \apj, 458, 524

\bibitem[\protect\citeauthoryear{{Izotov} \& {Thuan}}{{Izotov} \&
  {Thuan}}{2004}]{izotov04}
{Izotov} Y.~I.,  {Thuan} T.~X.,  2004, \apj, 602, 200

\bibitem[\protect\citeauthoryear{{Kewley}, {Dopita}, {Sutherland}, {Heisler} \&
  {Trevena}}{{Kewley} et~al.}{2001}]{kewley01}
{Kewley} L.~J.,  {Dopita} M.~A.,  {Sutherland} R.~S.,  {Heisler} C.~A.,
  {Trevena} J.,  2001, \apj, 556, 121

\bibitem[\protect\citeauthoryear{{Larson}}{{Larson}}{1974}]{larson74}
{Larson} R.~B.,  1974, \mnras, 169, 229

\bibitem[\protect\citeauthoryear{{Mac Low} \& {Ferrara}}{{Mac Low} \&
  {Ferrara}}{1999}]{maclow99}
{Mac Low} M.-M.,  {Ferrara} A.,  1999, \apj, 513, 142

\bibitem[\protect\citeauthoryear{{Marlowe}, {Heckman}, {Wyse} \&
  {Schommer}}{{Marlowe} et~al.}{1995}]{marlowe95}
{Marlowe} A.~T.,  {Heckman} T.~M.,  {Wyse} R.~F.~G.,    {Schommer} R.,  1995,
  \apj, 438, 563

\bibitem[\protect\citeauthoryear{{Martin}}{{Martin}}{1998}]{martin98}
{Martin} C.~L.,  1998, \apj, 506, 222

\bibitem[\protect\citeauthoryear{{Mendez} \& {Esteban}}{{Mendez} \&
  {Esteban}}{1997}]{mendez97}
{Mendez} D.~I.,  {Esteban} C.,  1997, \apj, 488, 652

\bibitem[\protect\citeauthoryear{{Moll}, {Mengel}, {de Grijs}, {Smith} \&
  {Crowther}}{{Moll} et~al.}{2007}]{moll07}
{Moll} S.~L.,  {Mengel} S.,  {de Grijs} R.,  {Smith} L.~J.,    {Crowther}
  P.~A.,  2007, \mnras, 382, 1877

\bibitem[\protect\citeauthoryear{{Nagao}, {Maiolino} \& {Marconi}}{{Nagao}
  et~al.}{2006}]{nagao06}
{Nagao} T.,  {Maiolino} R.,    {Marconi} A.,  2006, \aap, 459, 85

\bibitem[\protect\citeauthoryear{{Oey}, {Dopita}, {Shields} \& {Smith}}{{Oey}
  et~al.}{2000}]{oey00}
{Oey} M.~S.,  {Dopita} M.~A.,  {Shields} J.~C.,    {Smith} R.~C.,  2000, \apjs,
  128, 511

\bibitem[\protect\citeauthoryear{{Ohyama}, {Taniguchi}, {Iye}, {Yoshida},
  {Sekiguchi}, {Takata}, {Saito}, {Kawabata} et~al.,}{{Ohyama}
  et~al.}{2002}]{ohyama02}
{Ohyama} Y.,  {Taniguchi} Y.,  {Iye} M.,  {Yoshida} M.,  {Sekiguchi} K.,
  {Takata} T.,  {Saito} Y.,  {Kawabata} K.~S.,    et~al., 2002, \pasj, 54, 891

\bibitem[\protect\citeauthoryear{{Osterbrock} \& {Ferland}}{{Osterbrock} \&
  {Ferland}}{2006}]{osterbrock06}
{Osterbrock} D.~E.,  {Ferland} G.~J.,  2006, {Astrophysics of gaseous nebulae
  and active galactic nuclei}.
University Science Books

\bibitem[\protect\citeauthoryear{{Pittard}, {Dyson}, {Falle} \&
  {Hartquist}}{{Pittard} et~al.}{2005}]{pittard05}
{Pittard} J.~M.,  {Dyson} J.~E.,  {Falle} S.~A.~E.~G.,    {Hartquist} T.~W.,
  2005, \mnras, 361, 1077

\bibitem[\protect\citeauthoryear{{Sawyer}}{{Sawyer}}{1997}]{sawyer97}
{Sawyer} D.,  1997, DensePak Users Manual

\bibitem[\protect\citeauthoryear{{Shopbell} \& {Bland-Hawthorn}}{{Shopbell} \&
  {Bland-Hawthorn}}{1998}]{shopbell98}
{Shopbell} P.~L.,  {Bland-Hawthorn} J.,  1998, \apj, 493, 129

\bibitem[\protect\citeauthoryear{{Sidoli}, {Smith} \& {Crowther}}{{Sidoli}
  et~al.}{2006}]{sidoli06}
{Sidoli} F.,  {Smith} L.~J.,    {Crowther} P.~A.,  2006, \mnras, 370, 799

\bibitem[\protect\citeauthoryear{{Silich} \& {Tenorio-Tagle}}{{Silich} \&
  {Tenorio-Tagle}}{2001}]{silich01}
{Silich} S.,  {Tenorio-Tagle} G.,  2001, \apj, 552, 91

\bibitem[\protect\citeauthoryear{{Slavin}, {Shull} \& {Begelman}}{{Slavin}
  et~al.}{1993}]{slavin93}
{Slavin} J.~D.,  {Shull} J.~M.,    {Begelman} M.~C.,  1993, \apj, 407, 83

\bibitem[\protect\citeauthoryear{{Stil} \& {Israel}}{{Stil} \&
  {Israel}}{2002}]{stil02}
{Stil} J.~M.,  {Israel} F.~P.,  2002, \aap, 392, 473

\bibitem[\protect\citeauthoryear{{van Dokkum}}{{van
  Dokkum}}{2001}]{vandokkum01}
{van Dokkum} P.~G.,  2001, \pasp, 113, 1420

\bibitem[\protect\citeauthoryear{{Veilleux} \& {Osterbrock}}{{Veilleux} \&
  {Osterbrock}}{1987}]{veilleux87}
{Veilleux} S.,  {Osterbrock} D.~E.,  1987, \apjs, 63, 295

\bibitem[\protect\citeauthoryear{{Westmoquette}, {Exter}, {Smith} \&
  {Gallagher}}{{Westmoquette} et~al.}{2007a}]{westm07a}
{Westmoquette} M.~S.,  {Exter} K.~M.,  {Smith} L.~J.,    {Gallagher} J.~S.,
  2007a, \mnras, 381, 894

\bibitem[\protect\citeauthoryear{{Westmoquette}, {Gallagher}, {Smith},
  {Trancho}, {Bastian} \& {Konstantopoulos}}{{Westmoquette}
  et~al.}{2009b}]{westm09b}
{Westmoquette} M.~S.,  {Gallagher} J.~S.,  {Smith} L.~J.,  {Trancho} G.,
  {Bastian} N.,    {Konstantopoulos} I.~S.,  2009b, \apj, in press

\bibitem[\protect\citeauthoryear{{Westmoquette}, {Smith} \&
  {Gallagher}}{{Westmoquette} et~al.}{2008}]{westm08}
{Westmoquette} M.~S.,  {Smith} L.~J.,    {Gallagher} J.~S.,  2008, \mnras, 383,
  864

\bibitem[\protect\citeauthoryear{{Westmoquette}, {Smith}, {Gallagher} \&
  {Exter}}{{Westmoquette} et~al.}{2007b}]{westm07b}
{Westmoquette} M.~S.,  {Smith} L.~J.,  {Gallagher} J.~S.,    {Exter} K.~M.,
  2007b, \mnras, 381, 913

\bibitem[\protect\citeauthoryear{{Westmoquette}, {Smith}, {Gallagher},
  {Trancho}, {Bastian} \& {Konstantopoulos}}{{Westmoquette}
  et~al.}{2009a}]{westm09a}
{Westmoquette} M.~S.,  {Smith} L.~J.,  {Gallagher} J.~S.,  {Trancho} G.,
  {Bastian} N.,    {Konstantopoulos} I.~S.,  2009a, \apj, 696, 192

\end{thebibliography}
\bsp

\clearpage

\clearpage



\label{lastpage}
\end{document}